# Dual theory of the superfluid-Bose glass transition in disordered Bose-Hubbard model in one and two dimensions

Igor F. Herbut[*]

Department of Physics and Astronomy, University of British Columbia, 6224 Agricultural Road, Vancouver B. C., Canada V6T 1Z1

**Abstract:** I study the zero-temperature phase transition between superfluid and insulating ground states of the Bose-Hubbard model in a random chemical potential and at large integer average number of particles per site. Duality transformation maps the pure Bose-Hubbard model onto the sine-Gordon theory in one dimension (1D), and onto the three dimensional Higgs electrodynamics in two dimensions (2D). In 1D the random chemical potential in the dual theory couples to the space derivative of the dual field, and appears as a random magnetic field along the imaginary time direction in 2D. I show that the transition from the superfluid state both in 1D and 2D is always controlled by the random critical point. This arises due to a coupling constant in the dual theory with replicas which becomes generated at large distances by the random chemical potential, and represents a relevant perturbation at the pure superfluid-Mott insulator fixed point. At large distances the dual theory in 1D becomes equivalent to the Haldane's macroscopic representation of disordered quantum fluid, where the generated term is identified with the random backscattering. In 2D the generated coupling corresponds to the random mass of the complex field which represents vortex loops. I calculate the critical exponents at the superfluid-Bose glass fixed point in 2D to be $\nu = 1.38$ and $z = 1.93$, and the universal conductivity at the transition $\sigma_c = 0.25 e_*^2/h$, using the one-loop field-theoretic renormalization group in fixed dimension.

# 1  Introduction

In recent years, a new paradigm in physics of condensed matter has emerged under the name of superconductor-insulator transition. Among the quantum many-particle systems which are believed to exhibit this type of transition between their ground states at zero temperature are $^4He$ in random media [1], thin superconducting films [2] and Josephson junction arrays [3]. At $T=0$, as some parameter in Hamiltonian is varied, such a system is expected to show either zero or infinite dc linear resistance, with the transition caused by purely quantum fluctuations which qualitatively alter the many-body ground state. Experimentally, the superconductor-insulator transition is manifested as a change from continuously increasing to sharply decreasing resistance of the system as temperature is lowered, in accordance with the notion that at $T=0$ the ground state is either an insulator or a superconductor.

It has been argued [4], [5] that as a model for superconductor-insulator transitions it suffices to consider a Hamiltonian for interacting bosons in random external potential. The basic assumption behind the idea is that the transition corresponds to the onset of phase coherence of the already preformed Cooper pairs, and not to the formation of the pairs themselves. The ground state of the bosons can then be either superfluid (SF) or insulating; the insulating state may arise due to repulsive interactions, in which case it is an incompressible Mott insulator (MI) with a gap, or from a combination of interaction and localization effects, which may result in the formation of the gapless Bose glass phase (BG). Particularly interesting is the SI transition in two dimensions (2D). At finite temperature 2D are the lower critical dimension for superconductivity with the superconducting transition being of Kosterlitz-Thouless type [6]. Also, for non-interacting particles 2D are the lower critical dimension for Anderson localization, with all single particle states weakly localized [7]. Arguably, the main source of interest in 2D superconductor-insulator transitions comes from the suggestion [8] that right at the quantum ($T=0$) critical point dc conductivity should be finite and universal, determined only by the universality class of the transition. This exotic possibility of normal but universal diffusion results from a delicate balance between localization and superfluid fluctuations, and is unique for 2D.

Although the experimental support for universality of the conductivity at the transition is still rather weak [2], the calculation of this and other critical quantities for different universality classes of dirty-boson systems in 2D is a fundamental and unsolved problem. Without disorder, the transition in the system of interacting bosons on a lattice is always between Mott insulator and superfluid phases. With integer average number of particles per site, MI-SF transition is in the universality class of the classical 3D XY model [9], [5]. The critical exponents in this case are well known to be $z=1$, $\nu \approx 0.667$, and the universal conductivity has been calculated by variety of methods, including $1/N$ and the Monte Carlo calculations [10], and the $\epsilon$-expansion [11]. With disorder present, the critical behavior at the superconductor-insulator transition in 2D is much less understood, due to apparent non-existence of the upper critical dimension where the theory would have a critical point at a weak coupling [5], [12]. In fact, even the very nature of the transition at commensurate boson densities has been a matter of debate: simple arguments [5] suggest that at weak disorder the transition into the superfluid phase proceeds from the Bose glass phase, while most of the numerical studies see a direct MI-SF transition [13], [14]. In absence of a controlled analytical approach to the problem with disorder, most of our knowledge of the critical behavior at the BG-SF transition originates from the numerical studies [15], [16], [17]. On the analytical side, large-N expansion [18], real space methods [19] and strong coupling expansion [20] have all been brought to bear. While these calculations give valuable informations on the phase diagram of the system, the critical exponents are difficult to extract and less reliable. In particular, the universal conductivity at the BG-SF transition has hitherto been calculated only numerically.

In the present paper I study the critical behavior at $T = 0$ superconductor-insulator transition in 1D and 2D systems of interacting lattice bosons in Gaussian random chemical potential and at commensurate densities within dual description of the problem [21] which focuses on topological defects in the many-body ground state. This approach, in which one looks at the destruction of the long-range order by proliferation of the defects, has proven invaluable for several problems in the past, the Kosterlitz-Thouless transition being a prime example. Let me first describe the main results for the simpler 1D system. By duality the disordered Bose-Hubbard model maps onto the sine-Gordon theory in which the random chemical potential couples linearly to the space derivative of the dual field. The elimination of the fast modes of the dual field performed perturbatively in the strength of the lattice potential always leads to exponential suppression of the small periodic potential by disorder, so that at some smaller cutoff the lattice potential in the theory scales to zero. Most importantly, while renormalizing to zero the periodic potential together with the random chemical potential generates an additional disorder term, so that at long length scales the dual theory is without the periodic potential term, but contains two different disorder terms. This long length scale effective theory is equivalent to the Haldane's macroscopic representation [22], [5] of the disordered quantum fluid in 1D, where the generated term arises from the backscattering of the random potential. Once it helped generate the important backscattering term in the action, the random chemical potential which corresponds to forward scattering in the long length scale theory becomes redundant and can be removed from the problem by an appropriate change of variables. Thus the 1D disordered Bose-Hubbard model at large length-scales is described by the Haldane's macroscopic theory for dirty bosons with no trace of the lattice potential. The transition from the superfluid state is even at infinitesimal disorder controlled by the random fixed point in 1D [23], [5], which is interpreted as a sign that the transition is between the superfluid and the Bose glass phases. I end the analysis of the 1D problem with the discussion of the shape of the BG-SF phase boundary in the interaction- disorder plane.

The lesson from the 1D example is that the transition from the superfluid state in the disordered system is always into the Bose glass, and that it is controlled by the fixed point which can be located easily in the space of coupling constants of the dual theory. This suggests the same line of attack at the 2D version of the problem. A short account of the results in 2D has already been published [24]. The dual theory for 2D dirty bosons at $T = 0$ and with short-range repulsion is considerably more complicated than its 1D counterpart, and has the form of 2+1D classical, anisotropic Higgs electrodynamics (Ginzburg-Landau superconductor) in random magnetic field, directed and correlated along one dimension. The complex Higgs field describes closed vortex-loops in space and imaginary time, which interact via long-range forces mediated by the gauge field, whose Maxwell term in the action describes the sound excitations in the parent superfluid state. Condensation of the vortex field implies destruction of the superfluid by appearance of infinitely large vortex-loops, which is also reflected in the induced Higgs mass of the gauge field. At the pure critical point which controls the MI-SF transition, weak anisotropy which arises from the underlying quantum nature of the problem is argued to be irrelevant. More interestingly, weak disorder in the form of random correlated magnetic field turns out to be precisely marginal, and I assume it is marginally irrelevant. Similarly as in the 1D case, the random chemical potential generates additional disorder which can be understood as the random "mass" of the vortex field, which is strongly relevant at the pure MI-SF fixed point. The flow at the critical surface eventually ends in the attractive random critical point, which I again interpret as the BG-SF fixed point.

The critical exponents and the universal resistivity at the BG-SF critical point in 2D are determined to the lowest order in fixed-point values of the coupling constants in the dual theory, using the field-theoretic renormalization group in fixed dimension [25]. The calculations are performed directly in 2+1 dimensions since the duality between the Bose-Hubbard Hamiltonian and the Higgs electrodynamics holds only for the specific dimension, and I also want to determine the universal conductivity at the critical point, which is again finite exclusively in 2+1D. The idea of this method is to express the universal quantities like the critical exponents or the universal conductivity as perturbative series in renormalized, instead of bare, coupling constants, so that they assume finite values at the transition, determined by the critical point of the theory. The lowest order results $z = 1.93$ and $\nu = 1.38$ obtained this way compare reasonably well with the numerical calculations [17], even though I do not have a truly small parameter. The conductivity is obtained using the Kubo formula for the response of the disorder field to an external potential. By duality of charges and vortices this way one directly calculates the boson resistivity, instead of the conductivity, as a series in critical coupling constants. Inverting the final result I obtain the universal conductivity $\sigma_c = 0.25(e_*^2/h)$ at the quantum SF-BG critical point, in rough agreement with the numerical results [16], [17], and apparently smaller then the value suggested by experiments. Subtleties involved in interpretation of the experimental measurements of $\sigma_c$ are briefly discussed.

The paper is organized in the following manner. In the next section I define the dirty-boson lattice model and use duality transformations to arrive at the lattice theory in density representation in general dimension. In section 3 I study the dual theory in 1D and discuss the renormalization scenario and the phase diagram. In section 4 the dual representation in 2D is derived. The one-loop renormalization of the continuum version of the dual theory is performed and the critical exponents and the conductivity at the BG-SF transition are calculated. In the last section the summary and the discussion of the main conclusions is given. Technical details are left for appendices.

## 2 Hamiltonian and duality transformation

I study the Hamiltonian for a system of charge-$e_*$ bosons with on site repulsion $U > 0$, written in number-phase representation (Josephson junction array Hamiltonian) [5], [17], [21]:

$$\hat{H} = \frac{U}{2}\sum_i \hat{n}_i^2 - \sum_i h_i \hat{n}_i - t\sum_{i,\hat{\nu}} cos(\hat{\phi}_{i+\hat{\nu}} - \hat{\phi}_i), \quad (1)$$

where index $i$ labels the sites of a 1D (2D) lattice, $\hat{\nu} = \hat{x}$ ($\hat{\nu} = \hat{x}, \hat{y}$) is a unit vector, $\hat{n}_i$ represents a deviation from a large *integer* average number of bosons per site, $h_i$ is a Gaussian random chemical potential with $\overline{h_i h_j} = w_h \delta_{i,j}$ and $\overline{h_i} = 0$, and $\hat{\phi}_i$ is the phase variable canonically conjugate to the number of bosons, $[\hat{\phi}_i, \hat{n}_j] = i\delta_{i,j}$. The Hamiltonian (1) is in the same universality class as the Bose-Hubbard model [26], [17]. I calculate the partition function of the system

$$Z = Tr \exp\{-\beta \hat{H}\}, \quad (2)$$

$\beta = 1/T$, in the basis of states that diagonalize the particle numbers: $1 = \sum_{\{n_i\}} |\{n_i\}\rangle\langle\{n_i\}|$, $\{n_i\} = \{n_1, ... n_{N_L}\}$, $-\infty < n_i < \infty$ and integer, where $N_L$ is the total number of lattice sites. Following the standard procedure for construction of the path-integral representation of the partition function [27], one arrives at the expression:

$$Z = \sum_{\{n_{i,\alpha}\}} \int_{-\pi}^{\pi} \prod_{i=1}^{N_L} \prod_{\alpha=1}^{N} d\phi_{i,\alpha} \exp -\{\frac{\epsilon U}{2}\sum_{i,\alpha} n_{i,\alpha}^2 - \epsilon \sum_{i,\alpha} h_i n_{i,\alpha} - \quad (3)$$
$$t\epsilon \sum_{i,\hat{\nu},\alpha} cos(\phi_{i+\hat{\nu},\alpha} - \phi_{i,\alpha}) + i\sum_{i,\alpha} \phi_{i,\alpha}(n_{i,\alpha+1} - n_{i,\alpha})\},$$

with the boundary conditions only on the number of bosons $n_{i,0} = n_{i,N}$, $\epsilon = \beta/N$, and the limit $\epsilon \to 0$ is assumed. At this point there are two ways to proceed: utilizing the Poisson summation formula one may perform the sums over numbers of bosons and arrive at the phase representation, from which it follows that without disorder the MI-SF transition at $T = 0$ is in the universality class of the XY model [9], [5]. Disorder in this formulation enters via an imaginary boundary term, reducing significantly the utility of the phase representation in that case. Nevertheless, the phase representation leads naturally to the field theory for the superfluid order parameter, studied extensively in the literature [12]. Alternatively, one can attempt to integrate the phases out to be left with the partition function in number-representation. To that end I use the Villain representation of the cosine-term in the last equation:

$$\exp\{\epsilon t \sum_{i,\hat{\nu},\alpha} cos(\phi_{i+\hat{\nu},\alpha} - \phi_{i,\alpha})\} \simeq \sum_{\{m_{i,\alpha,\hat{\nu}}\}} \exp\{-\sum_{i,\hat{\nu},\alpha} \frac{m_{i,\alpha,\hat{\nu}}^2}{2\epsilon t} + i\sum_{i,\hat{\nu},\alpha} m_{i,\alpha,\hat{\nu}}(\phi_{i+\hat{\nu},\alpha} - \phi_{i,\alpha})\}, \quad (4)$$

where $\vec{m}_{i,\alpha}$ are one-component (two component) integer valued currents in 1D (2D) at the sites of the two (three) dimensional lattice labelled by indices $(i, \alpha)$. The integration over the phases now can be performed exactly, and its effect is to produce the constraint on the currents and numbers of bosons:

$$\nabla_{\vec{r}} \cdot \vec{m}_{i,\alpha} + \nabla_{\tau} n_{i,\alpha} = 0, \tag{5}$$

where $\nabla_{\vec{r}}$ and $\nabla_{\tau}$ are the lattice gradients (finite differences) in space and imaginary time directions, respectively. The last equation may be interpreted as a continuity equation, and the two (three) component vector $\vec{M}_{i,\alpha} = (\vec{m}_{i,\alpha}, n_{i,\alpha})$ represents the conserved bosonic current in 1D (2D). The Eq. (5) ensures that only closed current loops in space and imaginary time contribute to the statistical sum. The partition function now assumes the form

$$Z = \sum_{\{\vec{M}_{i,\alpha}\}} \delta(\nabla \cdot \vec{M}) \exp -\{\frac{\epsilon U}{2} \sum_{i,\alpha} n_{i,\alpha}^2 - \epsilon \sum_{i,\alpha} h_i n_{i,\alpha} + \frac{1}{2\epsilon t} \sum_{i,\alpha} \vec{m}_{i,\alpha}^2\}. \tag{6}$$

Note that the action in the exponent is completely real. As it stands, the last form of the partition function is still difficult to study analytically due to discreteness of the variables, and is more amenable to numerical methods [17]. It is possible however to further transform the partition function into a more telling expression [28], [29], [30], as in this context was pointed out by Fisher and Lee [21]. The form of the transformation differs in 1D and 2D, so I will address the two cases separately.

## 3 One dimension

### 3.1 Dual theory

In 1D the constraint $\nabla \cdot \vec{M} = 0$ may be resolved by introducing a *unique* integer scalar variable $N$, so that $\vec{M} = (\nabla_{\tau} N, -\nabla_x N)$. The theory then takes the form

$$Z = \lim_{y \to 0} \int \prod_{i,\alpha} dA_{i,\alpha} \exp -\{\frac{\epsilon U}{2} \sum_{i,\alpha} (\nabla_x A_{i,\alpha})^2 + \frac{1}{2\epsilon t} \sum_{i,\alpha} (\nabla_{\tau} A_{i,\alpha})^2 \tag{7}$$

$$-\epsilon \sum_{i,\alpha} h_i (\nabla_x A_{i,\alpha}) - \frac{1}{y} \sum_{i,\alpha} \cos(2\pi A_{i,\alpha})\},$$

where the limit $y \to 0$ serves to force the real-valued fields $\{A_{i,\alpha}\}$ to take only integer values. Hereafter I will take $T = 0$. Assuming that softening this constraint will not change the universality class of the transition [29], [30], in the continuum limit $a \to 0$, $\epsilon \to 0$ the dual theory becomes:

$$S = K \int dx d\tau [(\partial_{\tau} \theta(x,\tau))^2 + c^2 (\partial_x \theta(x,\tau))^2] - \int dx d\tau h(x) \partial_x \theta(x,\tau) \tag{8}$$

$$-v \int dx d\tau \cos 2\theta(x,\tau),$$

where $\pi A(x, \tau) = \theta(x, \tau)$, $c^2 = Uta^2$, $K = 1/(2t\pi^2 a)$, and $v$ may be understood as a (finite) strength of the lattice potential with the lattice constant $a$. Averaging over the Gaussian random potential by introducing replicas in standard way [31], one finally arrives at the form

$$S = K \sum_{\alpha=1}^{N} \int dx d\tau [(\partial_\tau \theta_\alpha(x,\tau))^2 + c^2(\partial_x \theta_\alpha(x,\tau))^2] - \quad (9)$$

$$w_h \sum_{\alpha,\beta=1}^{N} \int dx d\tau d\tau' \partial_x \theta_\alpha(x,\tau) \partial_x \theta_\beta(x,\tau') - v \sum_{\alpha=1}^{N} \int dx d\tau \cos 2\theta_\alpha(x,\tau),$$

where Greek indices now enumerate replicas, and the limit $N \to 0$ at the end of the calculation is assumed. If $w_h = 0$, dual to the Bose-Hubbard model (1) in 1D is the familiar sine-Gordon theory which has the transition in the XY universality class as $K$ or $v$ are varied [6]. At large $K$ the configurations with $\theta$ close to constant dominate the statistical sum, which implies a sharp particle number at each site and therefore disordered phases. Notice that by simple power counting the disorder coupling $w_h$ is relevant at the pure fixed point (located at $K^* c^* = 1/4$, $v^* = 0$, see below), so one may expect that with disorder the nature of the transition in the system should be changed. I will demonstrate that indeed that is the case, although, interestingly, the coupling $w_h$ will ultimately be irrelevant for the critical behavior. It is however crucial for bringing about the correct form of the effective theory at the large distances, to which I turn next.

## 3.2 Renormalization

I now integrate out the components of the field $\theta$ in Eq. (9) with the momenta $\Lambda/s < |k| < \Lambda$, $\Lambda \sim 1/a$ and $s \approx 1$, and with any frequency $\omega$, perturbatively in strength of the periodic potential $v$. The calculation is standard, with the only non-trivial step being a matrix inversion needed to obtain the propagator for the dual field $\theta$, which in presence of disorder is non-trivial in replica indices. The interested reader is refereed to the Appendix A for details, and here I only mention that this can be done exactly in the limit $N \to 0$. To the lowest order in $v$ the recursion relations are:

$$\frac{dv}{d\ln(s)} = (2 - \frac{1}{2Kc} - \frac{w_h}{2K^2 c^4})v + O(v^3), \quad (10)$$

$$\frac{dK}{d\ln(s)} = \frac{dc}{d\ln(s)} = O(v^2), \quad (11)$$

$$\frac{dw_h}{d\ln(s)} = w_h + O(v^2), \quad (12)$$

and I introduced the dimensionless combinations $w_h/\Lambda \to w_h$, $v/\Lambda^2 \to v$ and $\pi K \to K$. At $w_h = 0$ one may chose the units of lengths and energy so that $c = 1$, so the Eqs. (10)-(12) reduce to the celebrated Kosterlitz recursion relations [6] with $K$ playing the role of temperature for the equivalent classical 2D XY model. The crucial new feature brought by disorder is the presence of the $w_h$-term in the recursion relation (10) with a negative sign. Physically this is exactly what one expects: at large distances the random chemical potential washes away the periodic lattice potential. In fact, since $w_h$ is a relevant perturbation at the pure fixed point (at $v^* = 0$, $K^* = 1/4$), we see that it will make a weak enough periodic potential exponentially irrelevant at large distances for any $K$. If this would be the end of story we would come to a somewhat paradoxical conclusion that disorder always turns Mott insulator into a superfluid. This would follow from the observation that once $v = 0$ in (9) one can eliminate the $w_h$-term altogether by a shift in field $\theta(x, \tau)$ [23], no matter how large $w_h$ might have become at that scale. This would be incorrect however, for the following reason: to the second order in $v$ a new disorder term in the action (9) becomes *generated* during the mode elimination

$$-D(s) \sum_{\alpha,\beta=1}^{N} \int dx d\tau d\tau' \cos 2(\theta_\alpha(x,\tau) - \theta_\beta(x,\tau')), \tag{13}$$

with

$$D(s) = \frac{v^2 w_h \ln(s)}{4K^2 c^4 \Lambda^2}, \tag{14}$$

where $v$ and $w_h$ are now left dimensionful. The reader will recognize the term (13) as precisely the one appearing in Haldane's course-grained representation for the problem [22], [5], and which arises there from the components of the random potential with the wave vector $k \sim 2\pi\rho_0$ and may be understood as the backscattering term. The $w_h$-term would correspond to purely forward scattering in the large length scale theory in 1D. Once the coupling constant $D$ has become generated and small $v$ has scaled to zero we may freely shift the dual variables to completely eliminate the $w_h$-term from the problem. At a smaller cutoff $\Lambda'$ the effective theory therefore acquires the familiar form:

$$S_{\Lambda'} = K(\Lambda') \sum_\alpha \int dx d\tau [(\partial_\tau \theta(x,\tau))^2 + c(\Lambda')^2 (\partial_x \theta(x,\tau))^2] \tag{15}$$

$$- D(\Lambda') \sum_{\alpha,\beta} \int dx d\tau d\tau' \cos 2(\theta_\alpha(x,\tau) - \theta_\beta(x,\tau')),$$

with the effect of integration over the fast modes absorbed into the effective values of the coupling constants. For weak initial $v$ the value of $D(\Lambda')$ at the scale at which $v$ vanishes will be small, so one may monitor the change of the effective couplings in (15) during further mode elimination perturbatively in $D$. The recursion relations for this case are well known [23]:

$$\frac{dD}{d\ln(s)} = (3 - \frac{1}{K})D, \tag{16}$$

$$\frac{dK}{d\ln(s)} = \frac{4D}{K}, \tag{17}$$

where again I put $c = 1$. The transition from the superfluid state is thus always ultimately controlled by the random fixed point at $D^* = 0$ and $K^* = 1/3$, at which the correlation length exponent $\nu = \infty$ and the dynamical exponent is $z = 1$. This is in accord with the ref. [5] where the same conclusion was reached using only slightly different reasoning. Similar arguments have also recently been advanced in [32]. The random fixed point has a natural interpretation as the BG-SF critical point, although the nature of the insulating phase can not be directly inferred from the recursion relations at small $D$. Although the exponents have the same values as at the MI-SF fixed point in 1D, it is a different critical point: for instance, the superfluid two-point correlation function at the transition decays algebraically at large distances with the power 1/3, instead of 1/4 at the MI-SF transition [23].

## 3.3 The phase diagram

Although the precise phase diagram for the 1D Bose-Hubbard model will depend on the microscopic details, one can still understand some of its general features on the basis of the above renormalization scenario. Let me first assume that the periodic potential is weak, that $c = 1$, and think of the phases in $K - w_h$ plane. In these units $K = \sqrt{U/t}/2\pi^2$. At zero disorder, for $K < K_c \approx 1/4$ the system is in the superfluid phase, and for $K > K_c$ it is a Mott insulator. Now take the point with $K = K_c$, but $w_h > 0$ and small. According to the renormalization scenario discussed above, at some larger length scale the periodic potential disappears, and if it was weak initially we end up with an effective theory (15) at the smaller cutoff $\Lambda'$ with small $D(\Lambda')$ and $K(\Lambda') \approx 1/4$. The recursion relations (16)-(17) then imply that at even larger length scales $D$ will renormalize to zero, so the system is in the superfluid phase. The same remains to be true at small $w_h$ as long as $K < K_c(w_h) < 1/3$, where $K_c(w_h)$ is the initial value of $K$ for which $K(\Lambda') = 1/3$. Since $K$ increases under renormalization (i. e. the $O(v^2)$-term in Eq. (11) is positive), $K_c < K_c(w_h) < 1/3$ for a small $w_h$. I conclude that: 1) there is an intervening Bose glass phase between the superfluid and the Mott insulator at weak disorder, and 2) the BG-SF phase boundary must curve upward not to intersect $K = 1/3$ line (see Fig. 1).

What happens at large disorder $w_h$ can not be concluded on the basis of the perturbative renormalization group, since this would eventually correspond to large $D$ in the eqs. (16)-(17), which is outside of their domain of validity. I still expect that at any K at arbitrarily large disorder the system must be in the localized phase, so the BG-SF phase boundary should turn left, as suggested by the dashed line at Fig.1. What happens at smaller $K$ is an interesting question, since one is approaching a singular limit of non-interacting particles ($K = 0$). Since in this limit the system should be in localized phase for any disorder, I expect the phase boundary eventually to turn downwards, so that there may be a reentrance of the localized phase as $K$ decreases at fixed disorder. At the present time however this remains a speculation.

The reader could object that the conclusion that at any disorder the transition from the superfluid state is always into a Bose glass may be an artifact of the assumption that the disorder distribution is Gaussian and hence unbounded. One can easily show however that averaging over a bounded distribution (square shaped, for instance) again gives the $w_h$-term in (9) as the most relevant one, together with terms of higher order in $\theta$. Thus I still expect the dual theory at large length scales to have the same form as in Eq. (15), which leads to BG-SF transition. What will change is that a bounded distribution should lead to a finite region with the Mott insulator phase, as drawn in Fig. 1, instead of only the half-line at $w_h = 0$ and $K > K_c$, as in the Gaussian case. The transition between the Mott insulator and the Bose glass should be first order, since it corresponds to local collapse of the gap [20].

Finally, there is a question of what happens if at the microscopic scale $v \gg w_h$. At $K = K_c < 1/4$, without disorder one is right on the MI-SF separatrix in the $K - v$ renormalization group flow diagram. At small $w_h$ and $K = K_c$ it seems plausible than small disorder will make $v$ renormalize downward *faster* than it would without it, so that the system would scale towards small $v$ and some $K < 1/4$. If this is the case, all of the previous conclusions would remain the same. Thus even in this case it is natural to expect that the transition is again of BG-SF type.

# 4  Two dimensions

## 4.1  Dual theory

In 2+1-dimensional lattice theory in Eq. 6 the constraint on $\vec{M}$ may be resolved by introducing another set of integer three component vectors by $\vec{M} = \nabla \times \vec{N}$, where again the lattice version of the 3D curl operator is assumed. Then:

$$Z = \lim_{y \to 0} \sum_{\{\vec{l}_{i,\alpha}\}} \int_{-\infty}^{\infty} \prod_{i,\alpha} d\vec{A}_{i,\alpha} \exp\{i 2\pi \sum_{i,\alpha} \vec{l}_{i,\alpha} \cdot \vec{A}_{i,\alpha} + \frac{y}{2} \sum_{i,\alpha} \vec{l}_{i,\alpha}^2 - S[\nabla \times \vec{A}]\}, \qquad (18)$$

where,

$$S = \frac{\epsilon U}{2} \sum_{i,\alpha} (\nabla \times \vec{A})_\tau^2 - \epsilon \sum_{i,\alpha} h_i (\nabla \times \vec{A})_\tau + \frac{1}{2\epsilon t} \sum_{i,\alpha} (\nabla \times \vec{A})_{\vec{r}}^2, \qquad (19)$$

and I introduced an auxiliary set of integer three-component vectors $\{\vec{l}_{i,\alpha}\}$, the summation over which in the limit $y \to 0$ forces the real valued gauge-fields $\{\vec{A}_{i,\alpha}\}$ to take strictly integer values. To assure the gauge-invariance of the above expression the constraint $\nabla \cdot \vec{l} = 0$ must be enforced. This is the crucial difference from the 1D case, where one could have also introduced the two-component vectors $\{\vec{l}_{i,\alpha}\}$, but due to the lack of gauge freedom in 1+1 dimensions, this time without the constraint. So in 1D one may sum over $\{\vec{l}_{i\alpha}\}$ freely and end up with the theory (7). In 2+1D however we need new degrees of freedom to enforce the constraint. In fixed gauge $\nabla \cdot \vec{A} = 0$ the partition function can finally be written as

$$Z = \lim_{y \to 0} \int \prod_n [d\theta_n d\vec{A}_n] \exp\{\frac{1}{y} \sum_{n,\hat{\mu}} cos(\theta_{n+\hat{\mu}} - \theta_n - 2\pi \vec{A}_{n,\hat{\mu}}) - S[\nabla \times \vec{A}]\}, \qquad (20)$$

as can be checked by going backwards from Eq. (20) in Villain approximation and by performing the integrations over angles $\{\theta_n\}$, which enforces the constraint $\nabla \cdot \vec{l} = 0$. Index $n$ now labels the sites of a *three dimensional* lattice, $\hat{\mu} = \hat{x}, \hat{y}, \hat{\tau}$, and $-\pi < \theta_n < \pi$. With $\epsilon$ finite, one recognizes the last form of the partition function at $T = 0$ as the 3D "frozen" [28] anisotropic lattice superconductor in random magnetic field along one (time) axis.

The expression (20) is the sought dual representation of the original problem in 2D. The angles $\{\theta_i\}$ represent disorder variables [33], since they randomize the configurations of the gauge-field $\vec{A}$. In fact, without the coupling to the disorder variables, the action in Eq. (20) would be quadratic in the gauge-field, which may be identified with the sound excitation mode in the parent superfluid state of the original bosons. The long-range order for $\theta$-variables destroys superfluidity by gapping this sound mode in the pure case. In this sense the system described by Eq. (20) is dual to the original Hamiltonian (1): the superfluid phase of bosons corresponds to the disordered phase of $\theta$-variables, while the insulating phase (MI or BG) of bosons corresponds to the ordered phase of $\theta$-variables. Assuming that by leaving small but finite $y$ in the Eq. (20) the universality class of the transition at $T = 0$ is not changed [29], [24], one may write the soft-spin version of the lattice gauge theory in (20) as [34]:

$$S = \int d^2\vec{r}dz[|(\nabla - iq\vec{A}(\vec{r}, z))\Psi(\vec{r}, z)|^2 + \mu^2|\Psi(\vec{r}, z)|^2 + \frac{\lambda}{2}|\Psi(\vec{r}, z)|^4 + \qquad (21)$$
$$\frac{1}{2}(\nabla \times \vec{A}(\vec{r}, z))^2 + \frac{\gamma}{2}(\nabla \times \vec{A}(\vec{r}, z))^2_{\hat{z}} + h(\vec{r})(\nabla \times \vec{A}(\vec{r}, z))_{\hat{z}}],$$

which is a Ginzburg-Landau theory for a strongly type-II superconductor (assuming $\lambda >> q^2$) in transverse gauge $\nabla \cdot \vec{A} = 0$, in the random magnetic field along the $z$-axis. Non-zero parameter $\gamma$ indicates the inherent anisotropy of the quantum problem. The transition is tuned by varying the mass $\mu$, and the condensation of the complex disorder field $\Psi$ signals the destruction of superfluidity.

## 4.2 Renormalization

In the field theory (21) disorder appears as the random magnetic field along the $z$-direction. This term now plays the role similar to the $w_h$-term in 1D. Note that the random magnetic field depends only on two (spatial) coordinates and is completely correlated in time, as typical for quantum problems with static disorder [35]. The quantity of interest is the disorder-averaged zero-temperature free energy $F = -\overline{\ln Z}$. Using the replica trick to average over disorder, one arrives at the action for interacting replicas of the system

$$S' = \sum_{\alpha=1}^{N} \int d^2\vec{r}dz\{|(\nabla - iq\vec{A}_\alpha)\Psi_\alpha|^2 + \mu^2|\Psi_\alpha|^2 + \frac{\lambda}{2}|\Psi_\alpha|^4 + \frac{1}{2}(\nabla \times \vec{A}_\alpha)^2 +$$
$$\frac{\gamma}{2}(\nabla \times \vec{A}_\alpha)^2_{\hat{z}} - \frac{w_h}{2}\sum_{\beta=1}^{N}\int dz'(\nabla \times \vec{A}_\alpha(\vec{r}, z))_{\hat{z}}(\nabla \times \vec{A}_\beta(\vec{r}, z'))_{\hat{z}}\}. \qquad (22)$$

Similarly to 1D, the effect of averaging over disorder is to introduce an off-diagonal (in replica indices) contribution to the gauge-field propagator, as determined by the last term in Eq. (22).

The point of transition in the dual theory is reached by tuning the renormalized mass of the disorder field $m$ to zero. Simple power counting tells that one needs to consider four dimensionless coupling constants: the dual charge $\hat{q}^2 = q^2/p$, the quartic term coupling $\hat{\lambda} = \lambda/p$, the anisotropy parameter $\gamma$ and the width of the random field distribution $\hat{w}_h = w_h/p$, where $p$ is an arbitrary infrared scale. However, little analysis shows that this is not enough, and to complete the theory one needs to include one more coupling constant. Consider the diagram in Fig. 2: since the gauge-field propagator has off-diagonal part in replica indices, this diagram generates the quartic term which couples different replicas even though this term is absent initially. This is analogous to the way the standard quartic term $\lambda$ would get generated by the fluctuations of the gauge-field in the pure theory. We already saw that a similar generation of a new disorder term happens in 1D theory as well. To complete the theory one therefore must add the following term to the action in Eq. (22):

$$S'' = -\frac{w}{2}\sum_{\alpha,\beta=1}^{n}\int d^2\vec{r}dzdz'|\Psi_\alpha(\vec{r},z)|^2|\Psi_\beta(\vec{r},z')|^2, \tag{23}$$

with $w > 0$. Note that this term has precisely the form which would arise from averaging over an independently random mass in (21), correlated along the $z$-direction and with the Gaussian distribution with a width $w$. Thus the full theory $S = S' + S''$ contains five coupling constant at the critical surface $m = 0$: the fifth dimensionless coupling is $\hat{w} = w/\pi p^2$.

To address the quantum ($T = 0$) critical behavior at the SI transition I consider the renormalized theory right at the transition point $m = 0$:

$$S_r = \sum_{\alpha=1}^{N}\int d^2\vec{r}dz\{Z_\Psi|Z_{\hat{\mu}}(\nabla_{\hat{\mu}} - iq\vec{A}_{\alpha,\hat{\mu}})\Psi_\alpha|^2 + \frac{\lambda'}{2}|\Psi_\alpha|^4 - \frac{w'}{2}\sum_{\beta=1}^{N}\int dz'|\Psi_\alpha(\vec{r},z)|^2|\Psi_\beta(\vec{r},z')|^2$$
$$+\frac{Z_A}{2}(\nabla\times\vec{A}_\alpha)^2 + \frac{\gamma Z_\gamma}{2}(\nabla\times\vec{A}_\alpha)_{\hat{z}}^2 - \frac{Z_h w_h}{2}\sum_{\beta=1}^{N}\int dz'(\nabla\times\vec{A}_\alpha(\vec{r},z))_{\hat{z}}(\nabla\times\vec{A}_\beta(\vec{r},z'))_{\hat{z}}\}. \tag{24}$$

The lowest order contributions to the renormalized quantities are determined by the diagrams at Fig. 3. The infrared divergences are regulated by evaluating quartic vertices at the usual symmetric point

$$\vec{k}_i\cdot\vec{k}_j = \frac{(4\delta_{i,j}-1)p^2}{4}, \tag{25}$$

with an additional condition $k_{1,z} = k_{2,z} = -k_{3,z} = -k_{4,z}$, necessary because the disorder is correlated along the $z$-axis, so that individual $k_z$ components, and not only their sum, are conserved in $w$-vertices. The polarization is evaluated at a finite momentum $c\cdot p$, where $c$ is a constant to be specified shortly. The result of the one-loop renormalizations is:

$$Z_\Psi = 1 - \frac{1}{4}\hat{q}^2 \tag{26}$$

$$Z_{\hat{z}} = 1 + \frac{1}{8}\hat{w} \tag{27}$$

$$Z_A = 1 + \frac{c}{16}\hat{q}^2 \qquad (28)$$

$$\hat{\lambda}' = \hat{\lambda} - \frac{2\sqrt{2}+1}{8}\hat{\lambda}^2 - \frac{1}{2\sqrt{2}}\hat{q}^4 + 2\hat{\lambda}\hat{w}(\sqrt{2}\ln(1+\sqrt{2}) + \frac{1}{\sqrt{6}}\ln(\sqrt{2}+\sqrt{3})), \qquad (29)$$

$$\hat{w}' = \hat{w} + 2\hat{w}^2(\frac{1}{\sqrt{2}}\ln(1+\sqrt{2}) + \frac{1}{\sqrt{6}}\ln(\sqrt{2}+\sqrt{3})) - \frac{1}{2\sqrt{2}}\hat{\lambda}\hat{w}, \qquad (30)$$

$$Z_\gamma = Z_h = Z_{\hat{x}} = Z_{\hat{y}} = 1. \qquad (31)$$

The non-trivial numerical factors in Eqs. (29)-(30) arise from integration performed directly in 3D and from the choice of the renormalization point (25).

Let me first discuss the renormalization of the anisotropy parameter $\gamma$ and the random-field disorder variable $\hat{w}_h$. The renormalized couplings $\gamma_r = \gamma Z_\gamma/Z_A$ and $\hat{w}_{h,r} = \hat{w}_h Z_h/Z_A$ obey the equations:

$$\frac{d\gamma_r}{dt} = -\gamma_r(\frac{c}{16}\hat{q}_r^2 + O(\hat{q}_r^4)) + O(\gamma_r^2), \qquad (32)$$

$$\frac{d\hat{w}_{h,r}}{dt} = \hat{w}_{h,r}(1 - \eta_A) + O(\hat{w}_{h,r}^2), \qquad (33)$$

where $t = -\ln(p)$ and

$$\eta_A = \frac{d\ln(Z_A)}{dt}, \qquad (34)$$

is the anomalous dimension of the gauge-field propagator. A small anisotropy appears to be irrelevant at any fixed point with a finite dual charge. I checked that this remains true when the term $O(\hat{q}^4)$ is included in Eq. (32). Note that anisotropy in the gauge-theory can not simply be rescaled out, as it would be possible in the XY model. Although the dual theory (21) should be strongly and not weakly anisotropic in the continuum limit $a \to 0$, $\epsilon \to 0$, I will assume that there are no stable anisotropic ($\gamma \neq 0$) fixed points in the theory. This would be in agreement with what is known for the classical scalar electrodynamics close to four dimensions for large number of complex field components [36]. I therefore set anisotropy $\gamma = 0$ hereafter.

The scaling equation for the renormalized dual charge $\hat{q}_r^2 = \hat{q}^2/Z_A$ may be written exactly as

$$\frac{d\hat{q}_r^2}{dt} = \hat{q}_r^2(1 - \eta_A), \qquad (35)$$

as follows from the definition of the anomalous dimension of the gauge-field propagator. Thus at any fixed point in the theory with a finite dual charge, $\eta_A = 1$ [37]. In particular, this implies exact vanishing of the linear term in Eq. (33) at the MI-SF fixed point of the pure system. The fact that the scaling dimension of the random-field disorder $\hat{w}_h$ is simply $1 - \eta_A$ follows from the observation that the transverse part of polarization is diagonal in replica indices to the lowest order in $w_h$. The term in Fig. 4, for example, does not contribute to $Z_h$ since it turns out to be completely longitudinal.

The fate of the random-field disorder $\hat{w}_h$ is determined by higher order terms in the Eq. (33). Marginality of $\hat{w}_h$ follows from the quantum nature of disorder: if disorder was correlated along slightly less than one dimension its scaling dimension would have been negative. Guided by the 1D problem, I will make a simplifying assumption that apart from generating the quartic term $w$, disorder coupling $w_h$ is irrelevant at the MI-SF fixed point. If this assumption turns out to be incorrect, the calculation presented here would be pertinent to the crossover regime towards a new critical point with $\hat{w}_h \neq 0$. However, since $\hat{w}_h$ can at worst be marginally relevant at the pure fixed point, and therefore can initially grow much slower than $\hat{w}$, we are assured of a region for a small initial $\hat{w}_h$ where ignoring its possible growth is a sensible approximation.

The scenario in 2D is thus similar to what we already encountered in 1D: initially disorder in the problem is represented by the random magnetic field $\hat{w}_h$, which I assume is marginally irrelevant at the MI-SF fixed point. Before renormalizing to zero $\hat{w}_h$ generates another disorder-like coupling $\hat{w}$ which is relevant and which grows. In the rest of the article I will then set $\hat{w}_h = 0$, and follow the renormalization of the remaining three coupling constants.

The one loop $\beta$-functions for the remaining coupling constants $\hat{q}_r^2$, $\hat{w}_r = \hat{w}'/Z_\Psi^2$ and $\hat{\lambda}_r = \hat{\lambda}'/Z_\Psi^2$ are:

$$\frac{d\hat{q}_r^2}{dt} = \hat{q}_r^2 - \frac{c}{16}\hat{q}_r^4, \tag{36}$$

$$\frac{d\hat{\lambda}_r}{dt} = \hat{\lambda}_r(1 + \frac{1}{2}\hat{q}_r^2 + 2(\sqrt{2}\ln(1+\sqrt{2}) + \frac{1}{\sqrt{6}}\ln(\sqrt{2}+\sqrt{3}))\hat{w}_r) - \frac{2\sqrt{2}+1}{8}\hat{\lambda}_r^2 - \frac{1}{2\sqrt{2}}\hat{q}_r^4, \tag{37}$$

$$\frac{d\hat{w}_r}{dt} = \hat{w}_r(2 + \frac{1}{2}\hat{q}_r^2 - \frac{1}{2\sqrt{2}}\hat{\lambda}_r) + 2(\frac{1}{\sqrt{2}}\ln(1+\sqrt{2}) + \frac{1}{\sqrt{6}}\ln(\sqrt{2}+\sqrt{3}))\hat{w}_r^2. \tag{38}$$

For $\hat{w}_r = 0$ these equations reduce to those studied previously in the context of critical behavior of superconductors [37]. In that case for a choice $c > 5.17$ there are four fixed points in the theory: Gaussian ($\hat{q}_r^2 = \hat{\lambda}_r = 0$) and 3D XY ($\hat{q}_r^2 = 0$, $\hat{\lambda}_{r,xy} = 2.09$), both unstable in direction of the dual charge, tricritical ($\hat{q}_r^2 = 16/c$, $\hat{\lambda}_r = \hat{\lambda}_-$), unstable in $\hat{\lambda}$-direction, and the MI-SF critical point ($\hat{q}_r^2 = 16/c$, $\hat{\lambda}_r = \hat{\lambda}_+$), believed to be of "inverted" XY type [29] (see the inset on Fig. 5). The Gaussian and the tricritical fixed points are connected by the straight separatrix which determines the tricritical initial value of the Ginzburg-Landau parameter $\kappa = \sqrt{\lambda/2q^2} = \kappa_c$ in the superconducting problem. The tricritical value of $\kappa$ has been estimated both analytically [38] and numerically [39], and we may tune the renormalization point to match that number: $\kappa_c = 0.42/\sqrt{2}$ requires $c = 20$. It is worth noting that for $\hat{w} = 0$ our one-loop approximation gives encouragingly good estimates of the critical exponents both at the unstable XY ($\nu = 0.63$) and at the stable, charged fixed point ($\nu = 0.61$, $\eta = -0.20$) [40].

A small disorder $\hat{w}$ is relevant at the MI-SF fixed point, as follows from the last $\beta$-function, and more generally may be expected on the basis of Harris criterion [41]. Both perturbative [37] and non-perturbative [38], [29], [30] calculations of the correlation length exponent at the pure critical point yield values $\nu < 1$, suggesting relevance of the correlated random-mass disorder. Starting from the strongly type-II region ($\lambda >> q^2$), the flows are attracted by the random critical point at $\hat{q}_c^2 = 0.8$, $\hat{w}_c = 3.71$ and $\hat{\lambda}_c = 29.72$, which I identify as BG-SF critical point (Figure 5). The BG-SF critical point exists for any choice of the parameterization by $c$, unlike the charged fixed points in the pure theory. This is similar to the disordered classical scalar electrodynamics close to 4D [42], where randomness restores the critical point in the theory. The flow towards BG-SF critical point is oscillatory, which should lead to oscillatory corrections to scaling.

## 4.3 Critical exponents

Having obtained the approximate renormalization group flow and the fixed points, I proceed to calculate the critical exponents. First, note that the correlation length exponent $\nu$ and the dynamical exponent $z$ may be assigned a purely thermodynamical meaning, since the BG-SF transition temperature $T_c \sim \delta^{z\nu}$ and the zero-temperature compressibility $\kappa \sim \delta^{\nu(d-z)}$. $\delta$ is a parameter measuring the distance from the BG-SF transition at $T = 0$. One may therefore calculate these exponents from any representation of the partition function. The correlation length critical exponent $\nu$ is naturally defined away from the critical surface $m = 0$, while the calculation here is performed precisely at the critical surface. Nevertheless, I may still obtain $\nu$ from the knowledge of yet another renormalization factor $Z_{\Psi^2}$ [25] that accounts for the renormalization of the $|\Psi|^2$ term in the theory (22), evaluated at the critical surface $m = 0$. I find:

$$Z_{\Psi^2} = 1 + \frac{\hat{\lambda}}{4} - \frac{\hat{w}}{2}, \tag{39}$$

to the lowest order in the coupling constants (Figure 3 a). In analogy with the thermal critical phenomena, one may define the exponent $\gamma$ as:

$$\gamma = 1 - \lim_{p \to 0} \frac{d \ln Z_{\Psi^2}}{d \ln p^2}, \tag{40}$$

so that the exponent $\nu$ then follows from the standard scaling relation:

$$\nu = \frac{\gamma}{2 - \eta_d}. \tag{41}$$

$\eta_d$ is the anomalous dimension of the vortex field $\Psi$

$$\eta_d = -\lim_{p \to 0} \frac{d \ln Z_\Psi}{d \ln p}. \tag{42}$$

Using the lowest order eqs. (26) and (39), expressing the bare coupling constants in terms of the renormalized ones, and taking the infrared limit $p \to 0$, the correlation length exponent becomes a series in fixed point values of the renormalized coupling constants. To the lowest order I obtain:

$$\nu = \frac{1}{2}(1 + \frac{\hat{\lambda}_c - \hat{q}_c^2 - 4\hat{w}_c}{8}) = 1.38. \tag{43}$$

Similarly, the dynamical critical exponent $z$ is defined by:

$$z^{-1} = 1 + \lim_{p \to 0} \frac{d \ln Z_{\hat{z}}}{d \ln p}. \tag{44}$$

From Eq. (27), to the lowest order in the critical coupling constants I find:

$$z = 1 + \frac{\hat{w}_c}{4} = 1.93. \tag{45}$$

At the BG-SF critical point $\nu > 1$, as expected [41]. The dynamical exponent is very close to two, which was conjectured to be exact [5] in 2D, based on the assumption that compressibility at the BG-SF transition stays finite. The values are also in good agreement with the Monte Carlo results of ref. [17] ($\nu = 0.9 \pm 0.1$ and $z = 2 \pm 0.1$) and the real-space study of ref. [19] ($\nu = 1.4$ and $z = 1.7$).

In contradistinction to the critical exponents $\nu$ and $z$, the standard exponent $\eta$ characterizes the behavior of the correlation function for the superfluid order parameter at the transition, and can not be straightforwardly obtained from the dual theory.

## 4.4 Universal conductivity

Since the calculation is performed right at the transition point (where the mass of the dual field $m = 0$), one may calculate the universal dc conductivity in similar way as the critical exponents. Before proceeding however, two important points need to be clarified. First, the dual theory describes the dirty-boson system at $T = 0$, and therefore the universal conductivity I will calculate corresponds to $T \to 0$, $\omega \to 0$ order of limits. The reader should note that the order of limits in typical experiments is reversed, which raises subtle issues about the actual dissipation mechanism tested by the experiments [43] to which I will return in the concluding section. Nevertheless, one may compute $T = 0$ critical conductivity as a fundamental characteristic of the BG-SF fixed point. The second point concerns my intention to obtain the conductivity of bosons from the dual theory for the vortex field. This question is intimately linked with the mechanism of dissipation in the quantum ($T = 0$) problem. As already mentioned, the dual theory describes the disappearance of superfluidity due to destruction of the gapless mode, which, as argued, can happen only through condensation of vortices. The physical picture is that on the insulating side of the transition vortex loops in 3D classical electrodynamics blow up, completely disordering the phase of the superfluid order parameter. The ground state of the Bose glass insulator is therefore a superfluid of free vortices and antivortices, which represent the 2D space projection of the 2+1D vortex loops, while the original bosons are localized by random potential. In contrast, in the superfluid ground state bosons are phase coherent while vortices and antivortices are tightly bound into pairs. Right at the transition it seems natural to assume that both bosons and vortices should be mobile which then leads to a finite metallic conductivity [10]. Consider the current of bosons

$$I_b = e_* \dot{N}_b \tag{46}$$

from one end of the sample to another; motion of vortices perpendicular to the direction of the boson current induces the voltage by Josephson relation:

$$V_b = \frac{h \dot{N}_v}{e_*}, \tag{47}$$

where $\dot{N}_v$ is the vortex flux. The boson resistance is therefore

$$R_b = \frac{\dot{N}_v}{\dot{N}_b}\frac{h}{e_*^2}. \qquad (48)$$

On the other hand, vortices "see" bosons exactly the same way as bosons see them [44]; writing the above relations for current and "voltage" of vortices instead of bosons we find that:

$$R_v = \frac{V_v}{I_v} = \frac{\dot{N}_b}{\dot{N}_v}\frac{h}{e_v^2}, \qquad (49)$$

where $e_v$ is an arbitrary "charge" of a vortex. Combining the last two equations I find

$$\sigma_b^{-1} = \frac{\sigma_v}{e_v^2/h}\frac{h}{e_*^2}, \qquad (50)$$

where $\sigma_b$ and $\sigma_v$ are boson and vortex conductivities, respectively, and I used the fact that conductivity equals conductance in two dimensions.

The last expression enables one to calculate resistivity of the bosons $\sigma_b^{-1}$ by using the Kubo formula for vortex conductivity $\sigma_v$. The derivation of the Kubo formula in the theory with replicas is presented in Appendix B. The simplest way to evaluate the integrals over the correlation functions in Eq. (71) is to use the scheme of dimensional regularization. Recalling that the couplings $\lambda$ and $q^2$ have dimensions of frequency, it follows that in the static limit $\omega \to 0$ to the lowest order in these coupling constants we need to calculate the integrals in Eq. (71) only at $\vec{k} = 0$, so that they will give no contribution to $\sigma_v$ to this order. To the lowest order $\sigma_v$ can then be written as ($\vec{k} = (\vec{0}, \omega)$):

$$\sigma_v = (\sigma_{v,0} + \sigma_{v,1})\frac{e_v^2}{\hbar} \qquad (51)$$

where

$$\sigma_{v,0} = \lim_{\omega \to 0}\frac{2}{\omega}\left(\int \frac{d^3\vec{q}}{(2\pi)^3}\frac{1}{q^2} - 2\int \frac{d^3\vec{q}}{(2\pi)^2}\frac{q_x^2}{q^2(\vec{q}+\vec{k})^2}\right) = \frac{1}{16} \qquad (52)$$

in dimensional regularization, in agreement with the result of Cha et al. [10]. $\sigma_{v,1}$ is given by the diagrams at Fig. 6:

$$\sigma_{v,1} = -\frac{8}{\omega}\int \frac{d^3\vec{q}}{(2\pi)^3}\frac{\Sigma(\vec{q})}{q^4(\vec{q}+\vec{k})^2} - \frac{4w}{\omega}\int \frac{d^3\vec{q}d^2\vec{p}_\perp}{(2\pi)^5}\frac{q_x p_x}{q^2(p_\perp^2+q_z^2)(\vec{q}+\vec{k})^2(p_\perp^2+(q_z+\omega)^2)} \qquad (53)$$

in the static limit $\omega \to 0$. The second term in $\sigma_{v,1}$ vanishes due to $q_x$ integration. The self-energy appearing in the first term is to be calculated to the lowest order in $w$ at the transition point $m = 0$. For small momenta it can be written as

$$\Sigma(\vec{q}) = \frac{w}{2\pi}\ln|q_z|, \qquad (54)$$

as follows from the result for the dynamical exponent $z$. Inserting the last expression into Eq. (53) and introducing Feynman parameters to integrate over $q_x$ and $q_y$ leads to

$$\sigma_{v,1} = \frac{I}{4\pi^3}\frac{w}{\omega^2}. \tag{55}$$

The integral in the numerator is:

$$I = \int_{-\infty}^{\infty} dq_z \int_0^1 dt \frac{(t-1)\ln|q_z|}{q_z^2 + 2tq_z + t} = 5.89. \tag{56}$$

Recognizing the combination $w/\pi\omega^2$ as the dimensionless coupling $\hat{w}$, the lowest order result for the dc vortex conductivity becomes:

$$\sigma_v = \left(\frac{\pi}{8} + \frac{I}{2\pi}\lim_{\omega\to 0}\hat{w}\right)\frac{e_v^2}{h}, \tag{57}$$

and therefore formally divergent due to the infrared (static) limit. Rewriting $\sigma_v$ as a series in the renormalized coupling $\hat{w}_r$ however, to the lowest order

$$\sigma_v = Z_\Psi^{-2}\left(\frac{\pi}{8} + \frac{I}{2\pi}\lim_{\omega\to 0}\hat{w}_r\right)\frac{e_v^2}{h}, \tag{58}$$

where I introduced the wave-function renormalization factor $Z_\Psi$ to account for the fact that $\hat{w}_r$ describes correlation functions of the rescaled vortex field. To the lowest order the boson dc resistivity is then determined by the critical coupling constants as:

$$\sigma_b^{-1} = \left(\frac{\pi}{8} + \frac{\pi}{16}\hat{q}_c^2 + \frac{I}{2\pi}\hat{w}_c\right)\frac{h}{e_*^2}. \tag{59}$$

Numerical values of the critical couplings as found from the one-loop $\beta$-functions eqs. (36)-(38) yield the estimate of the universal conductivity of bosons at the BG-SF transition in 2D:

$$\sigma_b = 0.25\frac{e_*^2}{h}. \tag{60}$$

## 5 Summary and discussion

In the present paper it was shown that the system of disordered interacting bosons at a commensurate density has a superfluid-insulator transition at zero temperature governed by the random fixed point, which was interpreted as a sign of a BG-SF transition. This critical point can be naturally identified within the dual theory for the transition which describes topological defects in the ground state. The coupling constant which represents the disorder relevant at the pure MI-SF fixed point is not initially present in the microscopic dual theory, but becomes generated at larger length scales by the microscopic disorder and the lattice potential. I discussed the phase diagram in 1D, and calculated approximately the critical exponents $\nu$ and $z$ and the universal conductivity at the BG-SF transition in 2D.

I argued that only in the dual theory the BG-SF critical point may become apparent. It is not obvious to which terms would the generated disorder in Eq. (13) correspond in the theory written in terms of the superfluid order parameter, for example. The situation is reminiscent of the Kosterlitz-Thouless transition in the 2D XY model: while in principle the information on the transition is contained in the Ginzburg-Landau-Wilson functional, its extraction is hopelessly difficult; on the other hand, by focusing directly on the behavior of vortices the problem of the critical behavior can be solved exactly. It appears the same is accomplished by the dual formulation of the dirty boson problem, at least in 1D where the random critical point is perturbatively accessible. In 2D the dual theory has a strong-coupling critical point, and one can obtain the critical quantities only approximately.

It is interesting how the relevant disorder coupling is generated by the random chemical potential and the lattice, although they are both ultimately irrelevant at the BG-SF fixed point. In 1D the recursion relations imply that weak lattice potential always renormalizes to zero, but that while doing so it conspires with random chemical potential to induce another disorder term in the theory. At some larger length-scale the initial disorder becomes redundant in 1D, and one is left with an effective theory in the familiar Haldane's form. In a similar way, in 2D the random magnetic field generates the random mass for the vortex field, which immediately becomes the most relevant perturbation at the pure fixed point and whose flow ultimately determines the critical behavior. I assumed, but could not prove, that in 2D the initial disorder in form of the random magnetic field also becomes irrelevant once it generated the important random mass.

The proposed phase diagram in 1D has one counterintuitive feature: there is a region for $K > K_c$ where by *increasing* disorder one turns a Bose glass insulator into a superfluid. This comes as a consequence of renormalization in the theory and I believe is a genuine effect. The phase diagram in Fig. 1 is in disagreement with the density-matrix renormalization group study of Pai et al. [13], who claimed the direct MI-SF transition at small disorder. This work has been criticized recently by Prokof'ev and Svistunov [45] who performed a Monte Carlo calculation to find the phase diagram in qualitative agreement with the one discussed here.

The performed renormalization group calculation of the critical exponents in 2D, although still approximate, has a slight advantage when compared with other ad hoc schemes in that it leaves room for a systematic improvement by going to higher orders in perturbation theory. Although this is possible in principle, it is rather complicated since there are three couplings to consider. The relatively good agreement between my results and those of other approaches is therefore somewhat fortuitous. The reader should also note that the values of the critical exponents are close to the results of other studies [17], [19] which are performed at *incommensurate* densities of bosons. In 1D it is easy to show that nothing in the conclusions is changed by allowing $\overline{h} \neq 0$, since by shifting the dual field one may always place $\overline{h}$ under the cosine term in Eq. (9), so it disappears together with the lattice potential at large length scales. In 2D incommensurability corresponds to non-zero magnetic field along one direction, the effect of which at the random fixed point is presently not clear. Physically one would expect the incommensurability should become irrelevant at the BG-SF critical point, although it would be desirable to have a concrete calculation in support of this claim [46].

The universal conductivity at the BG-SF critical point in Eq. (60) turns out to be somewhat larger than in the Monte Carlo calculations of Wallin et al. [17], and smaller than in Batrouni et al [16]. It is smaller that at the MI-SF transition [10], [11], which seems plausible. It is also smaller than what is found in most experiments, where $\sigma \sim (2e)^2/h$, and it has recently been proposed that this discrepancy may not be a coincidence. Damle and Sachdev [43] pointed out that at the critical point $\sigma$ becomes a non-trivial universal function of $\omega/T$, and that the experimentally relevant limit corresponds to behavior of this function at small values of its argument. In this regime the relevant dissipation mechanism is the random scattering of thermally created quasiparticles, rather that the coherent transport of the field-induced defects in the ground state, which would correspond to the opposite limit $\omega/T \to \infty$. The direct comparison of the conductivity calculated here with the present experimental data [2] is therefore probably not appropriate. I may still note that number in Eq. (60) is intriguingly close to recent measurement in thin Pb film [47]: $\sigma_c \approx 0.25(2e)^2/h$. The experiment which would measure the high-frequency conductivity at low temperatures would be closer to probing the quantum regime for which the result in Eq. (60) is directly relevant.

Finally, the kinship between the BG-SF critical points in 1D and 2D suggests that one could try to approach the strong-coupling critical point in 2D via an expansion around 1D where the BG-SF fixed point happens to be located at zero disorder. Although the apparent dissimilarity between the forms of dual theories (9) in 1D and (22) in 2D at first sight makes this idea unlikely to work, all that is needed is analyticity of the critical exponents as functions of the dimension for $1 < D < 2$, for such an expansion not to be impossible in principle. Some steps in this direction have already been taken in [48], and this will be a subject of a future publication [49].

This work has been supported by NSERC of Canada and Izaak Walton Killam foundation.

# 6   Appendix A: dual field propagator in 1D

From the dual theory (9) the inverse of the propagator for the field $\theta$ is:

$$\hat{G}^{-1} = K(\omega^2 + c^2 k^2)\hat{I} - wk^2\delta(\omega)\hat{M}, \tag{61}$$

where $\hat{G}$ is $N \times N$ matrix, $\hat{I}$ is a unit matrix, and $\hat{M}$ is also a $N \times N$ matrix with all elements equal to one. The second term is due to disorder, and the delta-function in frequency originates in the quantum nature of the problem: disorder is represented by a potential random in space, but completely correlated in imaginary time. By noticing that $\hat{M}^2 = N\hat{M}$ we may write

$$\hat{G} = \frac{1}{K(\omega^2 + c^2 k^2)}\hat{I}(1 + \sum_{n=1}^{\infty}(\frac{w_h\delta(\omega)}{Kc^2})^n N^{n-1}\hat{M}). \tag{62}$$

In the relevant limit of vanishing number of replicas $N \to 0$ the propagator therefore takes a simple form:

$$\hat{G} = \frac{1}{K(\omega^2 + c^2 k^2)}\hat{I} + \frac{w_h \delta(\omega)}{K^2 c^4 k^2}\hat{M}. \tag{63}$$

# 7    Appendix B: Kubo formula for disordered dual theory

We want to find the response of the system described by Eq. (22) to the external vector potential $\vec{a}$ which couples to the disorder field via minimal coupling. I work in units where the charge of disorder field $e_v$ which determines the strength of the coupling to the external probe (not to be confused with the running dual charge $q$) and $\hbar$ are both set to unity. With the external probe, the kinetic part of the action in Eq. (22) becomes:

$$S' = \sum_{\alpha=1}^{N} \int d^2\vec{r}\, dz |(\nabla - iq\vec{A}_\alpha + i\vec{a})\Psi_\alpha|^2, \tag{64}$$

with the rest of the action unchanged. Disorder-averaged current of the disorder field is then given by:

$$\overline{\langle \vec{j}(\vec{r},z) \rangle} = -\frac{\delta \overline{F[\vec{a}]}}{\delta \vec{a}(\vec{r},z)}|_{\vec{a}=0}. \tag{65}$$

Angular brackets denote quantum average over the ground state and overline stands for the average over disorder configurations. The vortex conductivity is defined by linear response to the vortex "electric field" as:

$$\overline{\langle \vec{j}(\vec{r},z) \rangle} = -\int d^2\vec{r}'\, dz'\, \sigma_v(\vec{r}-\vec{r}', z-z')\frac{\partial \vec{a}(\vec{r}', z')}{i\partial z'}. \tag{66}$$

Integrating by parts the last equation and going to Fourier space I find:

$$\sigma_v(\omega) = -\frac{1}{\omega}\int d^2\vec{r}\, dz\, \exp(i\omega z)\frac{\delta^2 \overline{F[a]}}{\delta a_x(\vec{r},z)\delta a_x(\vec{0},0)}|_{\vec{a}=0}, \tag{67}$$

where I chose $\vec{a}$ to be along $\hat{x}$ direction. Using the replica trick to write $\overline{F}$ and performing the differentiations in (67) I obtain:

$$\sigma_v(\omega) = \frac{1}{\omega}\lim_{N \to 0}\frac{1}{N}\{2\sum_\alpha^N \int_\Psi \exp(-S)|\Psi_\alpha(\vec{0},0)|^2 - \int_\Psi \int d^2\vec{r}\, dz\, \exp(i\omega z)\exp(-S)j_x(\vec{r},z)j_x(\vec{0},0)\} \tag{68}$$

where the current is given by the standard expression summed over replicas:

$$\vec{j} = \sum_\alpha^N (i\Psi_\alpha^* \nabla \Psi_\alpha - i\Psi_\alpha \nabla \Psi_\alpha^* + 2q|\Psi_\alpha|^2 \vec{A}_\alpha). \tag{69}$$

Since
$$\langle A \rangle = \frac{\int_\Psi \exp(-S) A}{\int_\Psi \exp(-S)} = \int_\Psi \exp(-S) A, \qquad (70)$$
in the limit $N \to 0$, inserting the above expression for the current, the last equation can be rewritten in the final form:

$$\sigma_v(\vec{k}) = \frac{2}{\omega} \frac{e_v^2}{\hbar} \lim_{N \to 0} \{ \int \frac{d^3\vec{q}}{(2\pi)^3} \langle \Psi_1^*(\vec{q}) \Psi_1(\vec{q}) \rangle - \frac{2}{N} V \sum_{\alpha,\beta}^N \int \frac{d^3\vec{q} d^3\vec{p}}{(2\pi)^6} q_x p_x \langle \Psi_\alpha^*(\vec{q}) \Psi_\alpha(\vec{q}+\vec{k}) \Psi_\beta^*(\vec{p}+\vec{k}) \Psi_\beta(\vec{p}) \rangle$$
$$- \frac{2}{N} q^2 V^2 \sum_{\alpha,\beta}^N \frac{d^3\vec{q}_1 d^3\vec{q}_2 d^3\vec{p}_1 d^3\vec{p}_2}{(2\pi)^{12}} \langle \Psi_\alpha^*(\vec{q}_1) \Psi_\alpha(\vec{q}_2) \Psi_\beta^*(\vec{p}_1) \Psi_\beta(\vec{p}_2) A_{\alpha,x}(\vec{k}+\vec{q}_1-\vec{q}_2) A_{\beta,x}(\vec{p}_1-\vec{p}_2-\vec{k}) \rangle \}, (71)$$

with $\vec{k} = (\vec{0}, \omega)$, $V$ is the 3D volume, and I restored the correct units. Apart from appearance of the replica indices, first two terms are the same as in Cha et al. [10]. Simple power-counting in the last expression implies that $\sigma_v$ does not scale with frequency, and is thus given by an universal constant, in its natural units $e_v^2/\hbar$. This conclusion remains true in the fully renormalized theory as well, since the current operator can not acquire anomalous dimension due to gauge invariance.

**Figure Captions:**

Figure 1: The phase diagram for the commensurate 1D Bose-Hubbard model in random chemical potential. $K = \sqrt{U/t}/2\pi^2$ and $w_h$ measures the strength of disorder. Dashed line represents a more speculative part of the BG-SF phase boundary. The BG-MI transition is expected to be first order. For further discussion see the text.

Figure 2: Diagram of the order $w_h^2 q^4$ which generates the quartic interaction $w$ between different replicas in 2D. Dashed line represents the gauge field propagator.

Figure 3: The lowest order contributions to the self energy a), $\lambda'$ b), $w'$ c) and $Z_A$ d). Dashed line represents the gauge field propagator, the wavy line stands for the interaction vertex $w$ and dots denote the interaction vertex $\lambda$.

Figure 4: Purely longitudinal contribution to the off-diagonal (in replica indices) polarization.

Figure 5: Renormalization group flow and the fixed points in $\hat{q}^2 = \hat{q}_c^2$ plane. The inset shows the flow in $\hat{w} = 0$ plane [37].

Figure 6: Contributions of the order $w$ to the vortex conductivity $\sigma_v$.

∗ Address after September 1998: Department of Physics, Dalhousie University, Halifax, Nova Scotia, Canada B3H 3J5

# References

[1] P. A. Crowell, F. W. van Keuls, and J. D. Reppy, Phys. Rev. Lett. **75**, 1106 (1995); Phys. Rev. B **55**, 12620 (1997).

[2] Y. Liu and A. M. Goldman, Mod. Phys. Lett. B **8**, 277 (1994) and references therein; J. M. Valles Jr., R. C. Dynes and J. P. Garno, Phys. Rev. Lett. **69**, 3567 (1992); A. Yazdani and A. Kapitulnik, Phys. Rev. Lett. **74**, 3037 (1995).

[3] H. S. J. van der Zant, W. J. Elion, L. J. Geerligs, and J. E. Mooij, Phys. Rev. B, **54**, 10081 (1996).

[4] M. Ma and P. A. Lee, Phys. Rev B **32**, 5658 (1985); M. Ma, B. I. Halperin and P. A. Lee, Phys. Rev. B **34**, 3136 (1986).

[5] M. P. A. Fisher, P. B. Weichman, G. Grinstein and D. S. Fisher, Phys. Rev. B **40**, 546 (1989).

[6] J. M. Kosterlitz and D. J. Thouless, J. Phys. C **6**, 1181 (1973); J. M. Kosterlitz, J. Phys. C **7**, 1046 (1974).

[7] E. Abrahams, P. W. Anderson, D. Liccardello and T. V. Ramakrishnan, Phys. Rev. Lett. **42**, 673 (1979).

[8] M. P. A. Fisher, G. Grinstein and S. M. Girvin, Phys. Rev. Lett. **64**, 587 (1990).

[9] S. Doniach, Phys. Rev. B **24**, 5063 (1981).

[10] M. Cha, M. P. A. Fisher, S. M. Girvin, M. Wallin and A. P. Young, Phys. Rev. B **44**, 6883 (1991).

[11] R. Fazio and D. Zappala, Phys. Rev. B **53**, R8883 (1996).

[12] P. B. Weichman and K. Kim, Phys. Rev. B **40**, 813 (1989); R. Mukhopadhyay and P. B. Weichman, Phys. Rev. Lett. **76**, 2977 (1996).

[13] W. Krauth, N. Trivedi and D. Ceperly, Phys. Rev. Lett. **67**, 2307 (1991); R. V. Pai, R. Pandit, H. R. Krishnamurthy and S. Ramasesha, Phys. Rev. Lett. **76**, 2937 (1996); J. Kisker and H. Rieger, Phys. Rev B **55**, R11981 (1997).

[14] See also, F. Pazmandi, G. T. Zimanyi and R. Scalettar, Phys. Rev. Lett. **75**, 1356 (1995); F. Pazmandi and G. T. Zimanyi, Phys. Rev. B **57**, 5044 (1998).

[15] K. J. Runge, Phys. Rev. B **45**, 13136 (1992).

[16] G. G. Batrouni, B. Larson, R. T. Scalettar, J. Tobochnik, and J. Wang, Phys. Rev. B **48**, 9628 (1994).

[17] M. Wallin, E. Sorensen, S. M. Girvin and A. P. Young, Phys. Rev. B **49**, 12115 (1994).

[18] Y. Tu and P. B. Weichman, Phys. Rev. Lett. **73**, 6 (1994); Y. B. Kim and X. G. Wen, Phys. Rev. B **49**, 4043 (1994).

[19] L. Zhang and M. Ma, Phys. Rev. B **45**, 4855 (1992); K. Singh and D. Rokshar, Phys. Rev. B **46**, 3002 (1992).

[20] J. Freericks and H. Monien, Phys. Rev. B **53**, 2691 (1996).

[21] M. P. A. Fisher and D. H. Lee, Phys. Rev. B **39**, 2756 1989.

[22] F. D. M. Haldane, Phys. Rev. Lett. **47**, 1840 (1981).

[23] T. Giamarchi and H. J. Shulz, Phys. Rev. B **37**, 325 (1988).

[24] I. F. Herbut, Phys. Rev. Lett. **79**, 3502 (1997).

[25] G. Parisi, J. Stat. Phys. **23**, 49 (1980); J. Zinn-Justin, *Quantum Field Theory and Critical Phenomena* (Oxford University Press, Oxford, 1993), Chapter 27.

[26] M. P. A. Fisher and G. Grinstein, Phys. Rev. Lett. **60**, 208 (1988).

[27] See, for example: J. Negele and H. Orland, *Quantum Many-Particle Systems*, (Addison-Wesley, Reading, MA, 1988).

[28] M. Peskin, Ann. Phys. **113**, 122 (1978); P. R. Thomas and M. Stone, Nucl. Phys. B **144**, 513 (1978).

[29] C. Dasgupta and B. I. Halperin, Phys. Rev. Lett. **47**, 1556 (1981).

[30] I. F. Herbut, J. Phys. A: Math. Gen. **30**, 423 (1997).

[31] S. Edwards and P. W. Anderson, J. Phys. F **5**, 965 (1975).

[32] B. Svistunov, Phys. Rev. B **54**, 16131 (1996).

[33] H. Kleinert, Gauge Fields in Condensed Matter, (Singapore, World Scientific), vol 1.

[34] More formally, the theory in Eq. (21) can be derived by introducing a Hubbard-Stratonovich field $\Psi$ to decouple the kinetic energy term in (20), and then integrating over the dual angles $\theta$. The action (21) is the result of expansion in powers of $\Psi$, and $\langle\Psi\rangle \sim \langle e^{i\theta}\rangle$.

[35] S. N. Dorogovtsev, Phys. Lett. A **76**, 169 (1980); D. Boyanovski and J. Cardy, Phys. Rev. B **26**, 154 (1982); I. D. Lawrie and V. V. Prudnikov, J. Phys. C **17**, 1655 (1984).

[36] T. C. Lubensky and J-H. Chen, Phys. Rev. B **17**, 366 (1978).

[37] I. F. Herbut and Z. Tešanović, Phys. Rev. Lett. **76**, 4588 (1996); *ibid.*, **78**, 980 (1997).

[38] B. Bergerhoff, F. Freire, D. F. Litim, S. Lola, and C. Wetterich, Phys. Rev. B **53**, 5734 (1996).

[39] J. Bartholomew, Phys. Rev. B **28**, 5378 (1983).

[40] The reader may be wondering about the role of the parameter $c$ on which the critical exponents, otherwise universal quantities, seem to depend. The point is that the coefficients of the perturbative series for the $\beta$-functions are non-universal, which implies the same for the exponents to any finite order of the perturbation theory in fixed dimension. This is an artifact of the perturbative calculation without any physical meaning. Our trick is to use this to our advantage and choose the scheme (parameterized by "c") which yields the result on $\kappa_c$ in agreement with the previous numerical calculations.

[41] A. B. Harris, J. Phys. C: Sol. St. Phys. **7**, 1671 (1974); J. T. Chayes, L. Chayes, D. S. Fisher, and T. Spencer, Phys. Rev. Lett. **57**, 2999 (1986).

[42] D. Boyanovsky and J. L. Cardy, Phys. Rev. B **25**, 7058 (1982).

[43] K. Damle and S. Sachdev, Phys. Rev. B **56**, 8714 (1997).

[44] F. D. M. Haldane, Phys. Rev. Lett. **51**, 605 (1983).

[45] N. V. Prokof'ev and B. V. Svistunov, preprint cond-mat/9706169, unpublished.

[46] See ref. 12 for an argument in favour of this proposition within the theory for the superfluid order parameter.

[47] K. Kagawa, K. Inagaki, and S. Tanda, Phys. Rev. B **53**, R2979 (1996).

[48] I. F. Herbut, Phys. Rev. B **57**, 1303 (1998)

[49] I. F. Herbut, cond-mat/9802141 (unpublished).

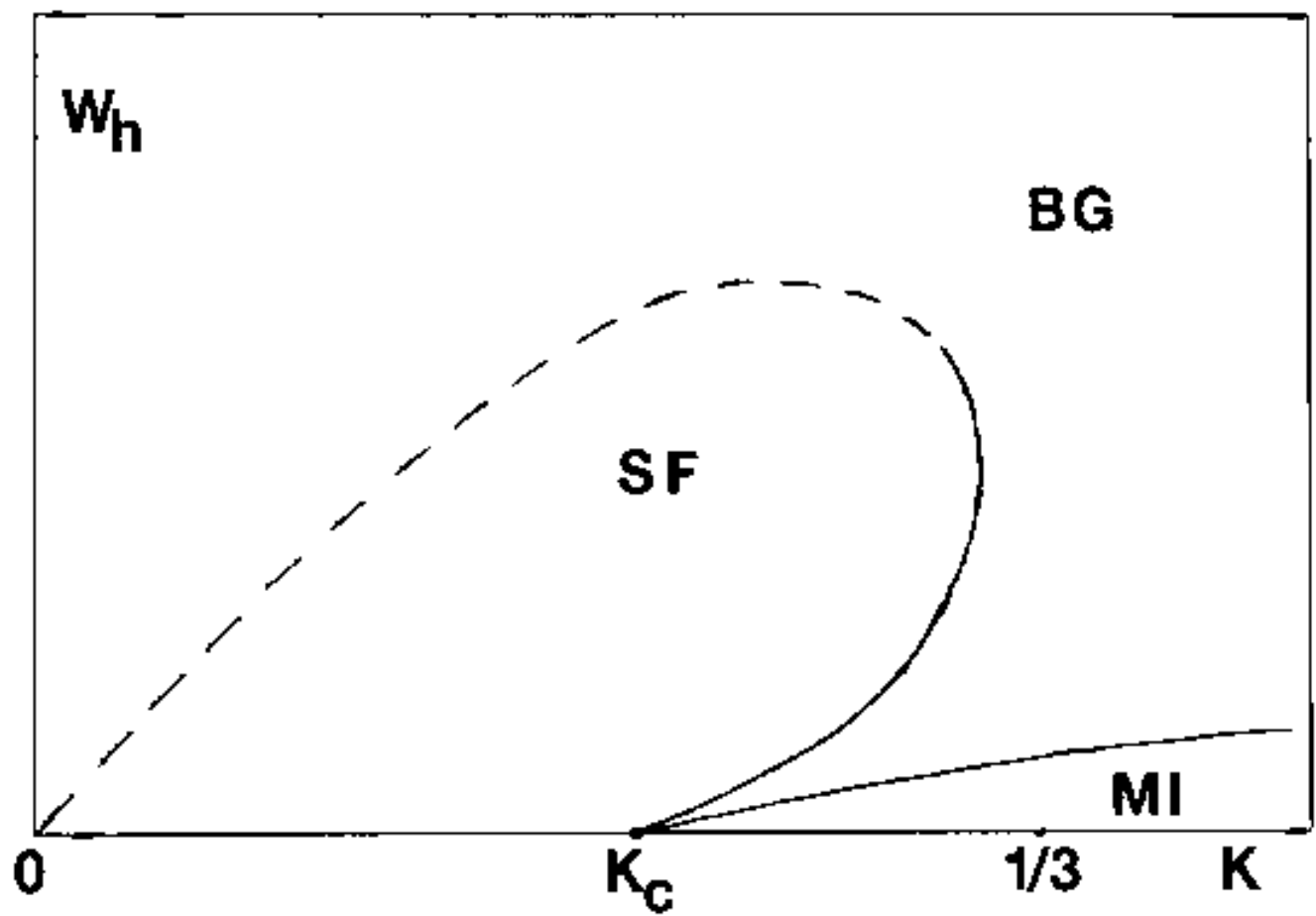

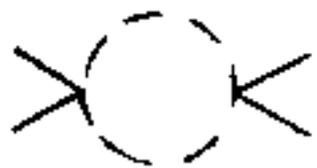

a

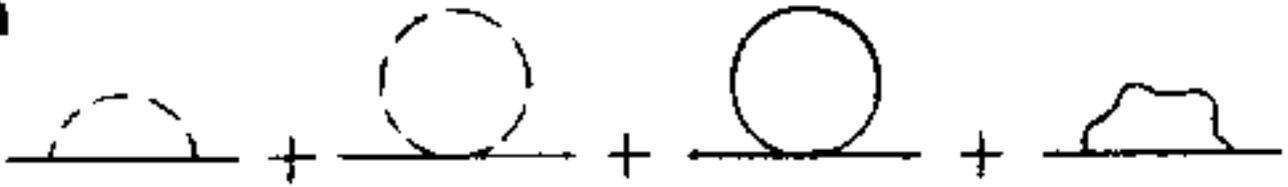

b

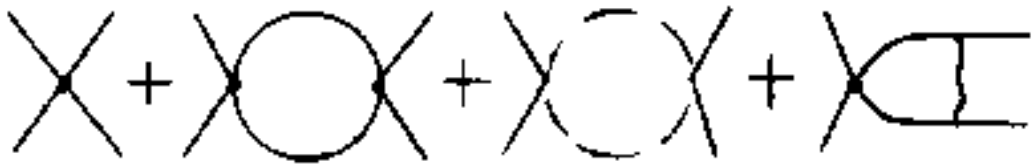

c

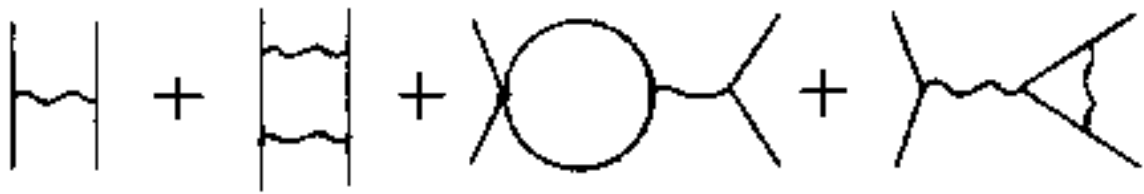

d

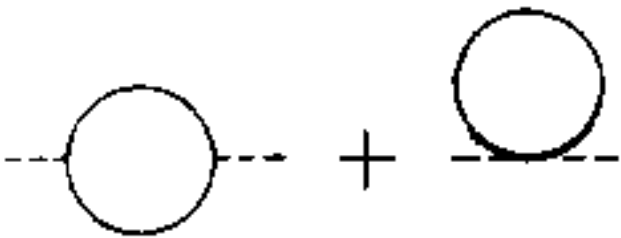

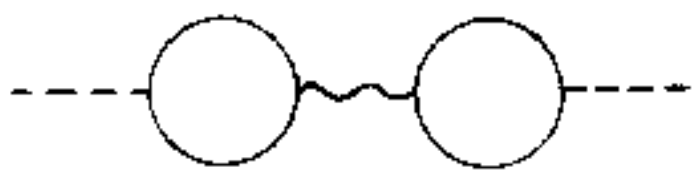

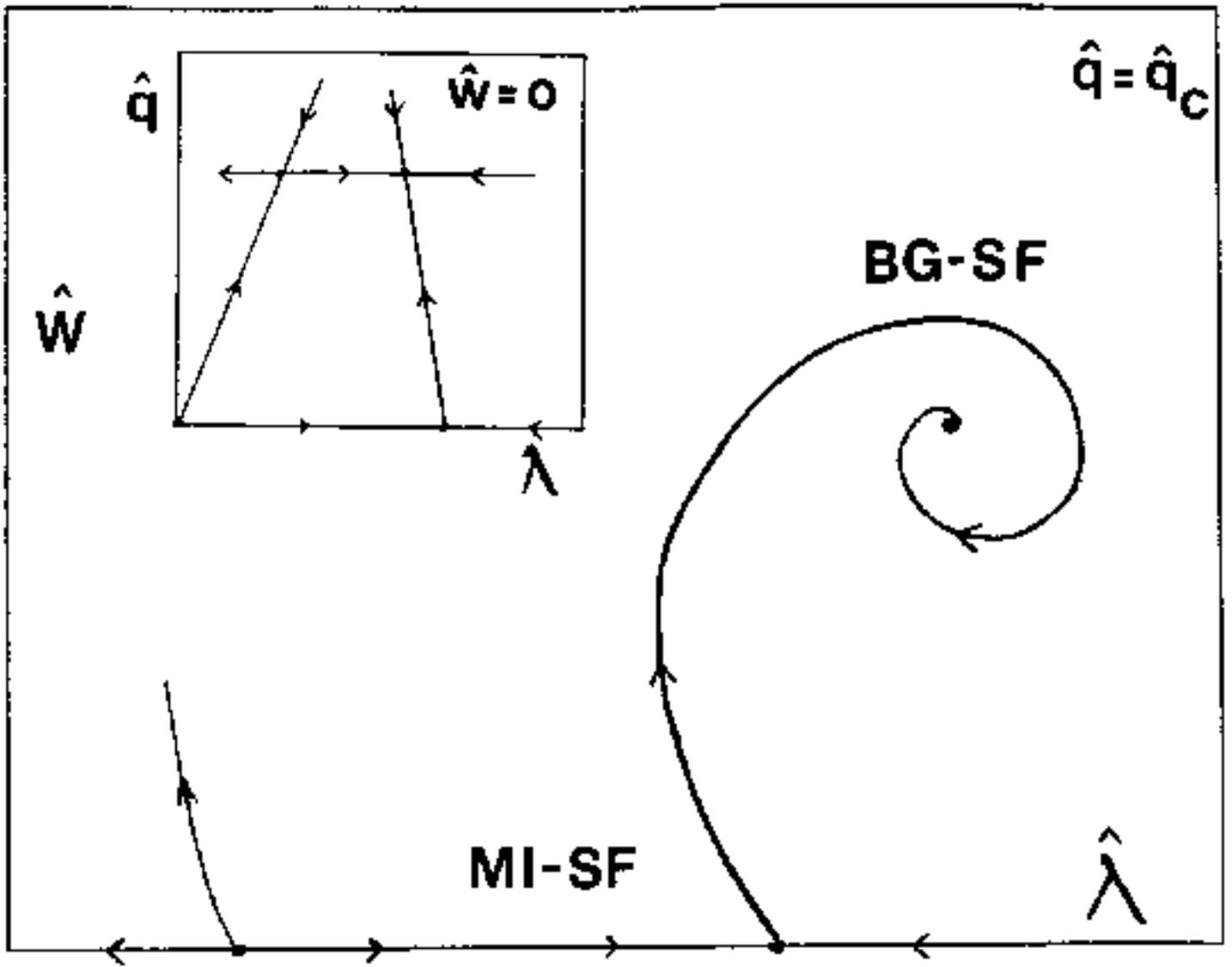

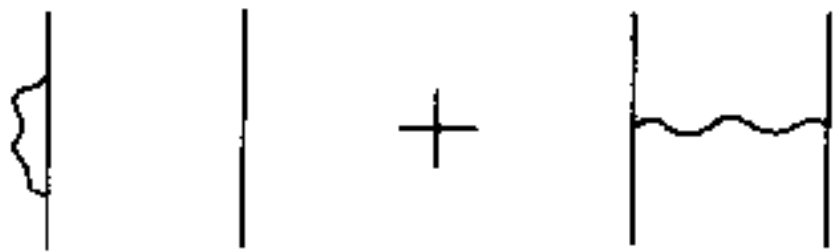